\def\spose#1{\hbox to 0pt{#1\hss}}
\def\approxlt{\mathrel{\spose{\lower 3pt\hbox{$\sim$}}
	\raise 2.0pt\hbox{$$<$$}}}
\def\approxgt{\mathrel{\spose{\lower 3pt\hbox{$\sim$}}
	\raise 2.0pt\hbox{$>$}}}

\def\multleft#1{\hbox to size{\vbox {\halign {\lft{##}\cr #1}}\hfill}\par}
\def\multright#1{\hbox to size{\vbox {\halign {\rt{##}\cr #1}}\hfill}\par}

\def\today{\ifcase\month\or January\or February\or March\or April\or May\or
      June\or July\or August\or September\or October\or November\or December\fi
      \space\number\day, \number\year}
\def\${\thinspace}

\def\boxit#1{\vbox{\hrule\hbox{\vrule\kern3pt\vbox{\kern3pt
          #1 \kern3pt}\kern3pt\vrule}\hrule}}

\documentstyle[psfig]{mn}
\begin{document}
\hsize=6truein
\renewcommand{\thefootnote}{\fnsymbol{footnote}}
\title{On the influence of resonant absorption on the 
iron emission line profiles from accreting black holes}
\author[Mateusz Ruszkowski and Andrew C. Fabian]
{\parbox[]{6.in} {Mateusz Ruszkowski and Andrew C. Fabian}
\\
\footnotesize
Institute of Astronomy, Madingley Road, Cambridge CB3 0HA \\}
\maketitle
\begin{abstract}  
The fluorescent iron K$\alpha$ emission line 
profile provides an excellent
probe of the innermost regions of active galactic nuclei. 
Fe XXV and Fe XXVI in diffuse plasma above the accretion disc
can affect the X-ray spectrum by iron K$\alpha$ resonant absorption.
This in turn can influence
the interpretation of the data and the estimation 
of the accretion disc and black hole parameters.
We embark on a fully relativistic computation of this effect
and calculate the iron line 
profile in the framework of a specific model in which rotating,
highly ionized and resonantly-absorbing plasma 
occurs close to the black hole. 
This can explain the features seen in the iron K$\alpha$ line profile 
recently obtained by Nandra et al. (1999) for the Seyfert 1 galaxy NGC 3516.
We show that the redshift of this feature can be mainly gravitational in origin
and accounted for without the need to invoke fast accretion of matter 
onto the black hole. New X-ray satellites such as XMM, ASTRO-E and Chandra 
provide excellent opportunities to test the model against high quality 
observational data.
\end{abstract}
\section{Introduction}
The fluorescent K$\alpha$ iron emission 
line has been observed in many active galactic nuclei and shown to possess
a broad and redshifted profile (Fabian et. al 1994, Mushotzky et al. 1995,
Tanaka et al. 1995, 
Reynolds 1997, Nandra et. al 1997). It has been demonstrated,
for example, 
in the best studied case of the Seyfert 1 galaxy MCG-6-30-15, that the strong 
continuum variability is accompanied by changes in 
the iron line itself (Iwasawa et al. 1996, Lee et al. 1999).
These results show that the 
fluorescent Fe K$\alpha$ line originates from the very central parts of the 
accretion disc close to the black hole, giving a  
natural explanation for both the rapid variability and the 
strong redshift detected in these objects. This opens a unique opportunity
constrain the spin of the black hole,
its mass and the structure of the accretion disc.\\
\indent
Recently Nandra et. al (1999)
reported on a long ASCA observation of the Seyfert 1 galaxy NGC 3516. Their
integrated iron line profile contains a factor of $\sim3.5$ times more photons
than that of MCG-6-30-15 and therefore at the present time 
provides an excellent 
opportunity to study the central parts of an active nucleus. 
Nandra et al. (1999) found tentative evidence for a redshifted
absorption feature superimposed on the emission line profile.
They interpret it as being consistent with 
resonant scattering in infalling material.
Motivated by such an interesting possibility, 
we embark on theoretical modelling of a similar effect.
The crucial ingredient of our model is a strong gravitational field and
therefore we fully account for all relevant general relativistic effects.
We assume that the black hole and the central parts of the accretion disc
are embedded in a cloud of hot, rapidly rotating and highly ionized
plasma in which iron is not totally stripped of all electrons.
Hydrogen and helium-like iron ions are then capable of
producing significant absorption and thus of changing the shape of the 
disc emission line profile. 
If confirmed by further observations, this effect would suggest that
future high quality data should be analyzed with 
the possibility of resonant absorption in mind,
as the presence of
optically-thin material surrounding the central region of an AGN could
modify the emission line profile and therefore affect the interpretation
of such data.
The presence of the absorption features can also
shed new light on the properties of the accretion flow near black holes,
as the position of the gravitationally redshifted absorption line 
will additionally depend on the velocity pattern close to the black hole.\\
\indent
The paper is organized as follows: the next section contains the
description of our approach to calculate the absorption features.
Section 3 is devoted to the general
presentation and discussion of our results and also contains
a discussion of the iron line shape  
in the context of the recently obtained iron line profiles of NGC 3516
by Nandra et. al (1999).
\section{Description of the method and basic assumptions}
We assume that the accretion disc is optically thick and geometrically thin
and that 
the primary source of X-ray radiation is a hot and optically thin corona
located just above the accretion disc. The central black hole and the 
accretion disc with the corona are embedded in a cooler 
($\sim 10^{7}- 10^{8}$K), extended cloud of plasma (see Fig. 1).
\begin{figure*}
 \centerline{\psfig{figure=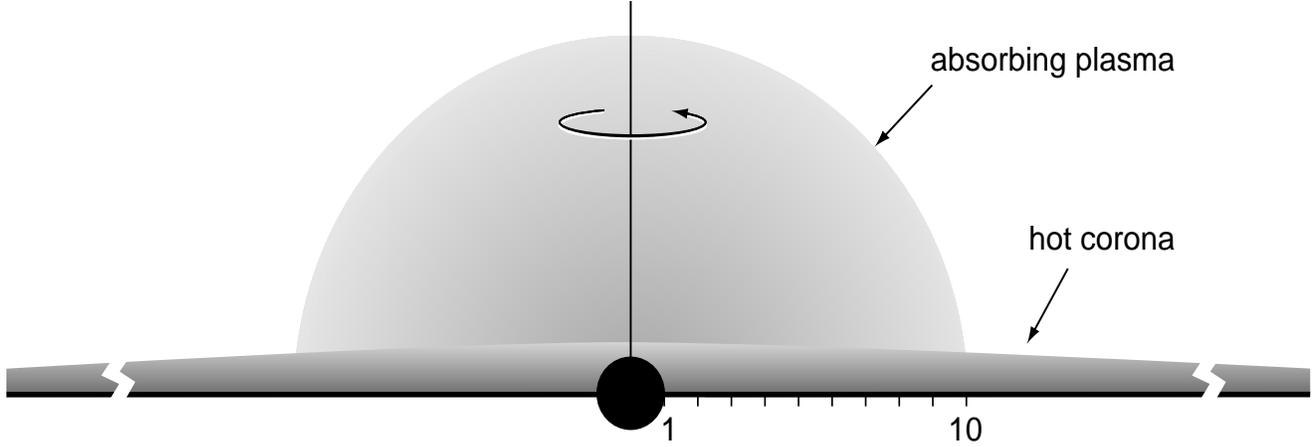,width=0.98\textwidth,height=0.25\textheight,angle=0}}
\caption{The assumed geometry of the central region}
\end{figure*}
The primary power law flux is incident on the disc and produces fluorescent 
photons. The fluorescent line and continuum photons then pass through
and may interact with such plasma. 
Under such conditions
a significant fraction of iron ions
are in the hydrogen (Fe XXVI) and helium-like (Fe XXV) states. These ions
can resonantly absorb the fluorescent line and continuum 
photons emerging from the accretion disc and corona.
The permitted transitions of interest are
$1s (^{2}S) - 2p (^{2}P)$ of H-like iron $(E_{\rm{H}}=6.9$ keV)
and $1s^{2} (^{1}S) - 1s2p (^{1}P)$ of He-like iron $(E_{\rm{He}}=6.7$ keV) 
with the large oscillator strengths $f_{\rm{lu}}$ equal 
0.416 for H-like and 0.794 for He-like (Kato 1976).
This means that absorption lines 
can potentially change 
the shape of the observed 
broad iron line profile which is generated in the accretion
disc with the rest energy $E_{\rm{disc}}=6.4$ keV. 
\subsection{The optical depth}
We now proceed to 
describe the calculation method 
for the optical depth of resonant absorption.
We first estimate the characteristic
distance $\lambda_{\rm{Sob}}$ over which the energy of a photon, 
as seen by an observer
who is at rest with the hot plasma, changes by the Doppler thermal width
$\Delta \nu_{\rm{D}}=(\nu_{\rm{abs}}/c)(2kT/A m_{\rm{p}})^{1/2}$, where 
$m_{\rm{p}}$ is a mass of
a proton, $A$ stands for the ion mass number and $\nu_{\rm{abs}}$ 
is the frequency
of the absorption line. We use the equation
(Novikov \& Thorne, 1973):
\begin{equation}
\left|\frac{dl}{d\nu}\right|=
\left(\left|\nu_{\rm{abs}}(p_{\mu}u^{\mu})_{;\alpha}p^{\alpha}\right|\right)^{-1}
\equiv\lambda=
\frac{\lambda_{\rm{Sob}}}{\Delta\nu_{\rm{D}}},
\end{equation}
where $l$ is the proper length as measured in the local rest frame of the
plasma flow.
This frame-independent equation is derived easily by geometrical arguments 
in the locally non-rotating frame. In the above equation $p_{\mu}$ is
the 4-momentum of a photon, $u^{\mu}$ is the 4-velocity of the absorbing
medium and $\nu_{\rm{abs}}$ is the rest frequency of the absorption line.
We choose the velocity field $u^{\mu}$ to be of the form:
\begin{equation}
{\bf u}=C(\partial_{\rm{t}}+\Omega\partial_{\phi}),
\end{equation}
where
\begin{equation}
C=(-g_{\rm{tt}}-2\Omega g_{\rm{t}\phi}-\Omega^{2}g_{\phi\phi})^{-1/2}
\end{equation}
and $\Omega=u^{\phi}/u^{\rm{t}}$ 
is the angular velocity of the absorbing medium as seen at 
infinity. Note that the choice of $\Omega$ cannot be arbitrary and 
certainly the angular velocity must be constrained by the causality condition:
$g_{\rm{tt}}+2\Omega_{\phi \rm{t}}+\Omega^{2}g_{\phi\phi}<0$.
For example, note that the assumption of constant specific angular momentum 
$l_{\rm{s}}\equiv -u_{\phi}/u_{\rm{t}}$
on spheres of fixed r in Boyer-Lindquist coordinates
(Kurpiewski \& Jaroszy\'nski 1999) violates this condition.
Therefore, in our simulations we adopt the following definition of $\Omega$:
\begin{equation}
\Omega=\left(\frac{\theta}{\pi/2}\right)\Omega_{\rm{K}}+
\left[1-\left(\frac{\theta}{\pi/2}\right)\right]\omega ,
\end{equation}
where $\theta$ is the poloidal Boyer-Lindquist coordinate, 
$\Omega_{\rm{K}}=(r^{3/2}+a)^{-1}$ is the Keplerian angular velocity and
$\omega$ is the angular velocity of the gravitational drag, i.e. we 
assume that the hot plasma follows the velocity of the disc close to its 
surface and gradually changes its angular velocity 
to reach the value $\omega$ on the polar axis
(note that at the black hole horizon radius $\Omega=\omega$ for all $\theta$).
From Eq. 1 we get:
\begin{equation}
\lambda_{\rm{Sob}}=\frac{1}{c}\left(\frac{2kT}{Am_{\rm{p}}}\right)^{1/2}\lambda=
5.7\times 10^{-4}\left(\frac{T}{10^{8}\rm{K}}\right)^{1/2}\lambda, 
\end{equation}
It has been checked numerically that for the overwhelming majority of 
trajecories connecting the accretion disc and the observer (for different
inclinations of the disc)
$\lambda_{\rm{Sob}}\ll\lambda_{\rm{c}}$, 
where $\lambda_{\rm{c}}$ is a characteristic length scale over which
the density remains roughly constant. 
However in our calculations we assume that the cloud has a uniform density 
distribution $n_{\rm{H}}$ in the local rest frame within a given radius $R_{c}$
(i.e. $\lambda_{\rm{c}}=R_{\rm{c}}$). 
In the real situation, the absorption feature is likely to be variable
(as indeed observed by Nandra et al. 1999) and the density distribution
may correspond to an unstable configuration with a clumpy structure.
Since we only intend to show what the temporal shape of the line might be, 
we argue that at the present stage 
it is not necessary to consider a more sophisticated density distribution
corresponding to a stable configuration.
The above constraints on $\lambda_{\rm{Sob}}$ ensure 
that we may use the Sobolev 
approximation to describe the resonant absorption in the plasma, i.e.
we may assume that the absorption occurs locally and 
that the medium is transparent for most 
of the trajectory of a photon with a given emission energy.
Thus in the further 
analysis we model the absorption line as a Dirac delta function. 
The optical depth per absorption event is then given by the formula
(compare with Castor 1970, Shu 1991):
\[
\tau=5.5\times10^{-2}\left(\frac{E_{\rm{abs}}}{6.7 \rm{keV}}\right)^{-1}
\left(\frac{A_{\rm{Fe}}}{2A_{\rm{Fe}\sun}}\right) 
\left(\frac{f_{\rm{lu}}}{0.5}\right)\times
\]
\begin{equation}
\hspace{1cm}\left(\frac{f_{\rm{l}}N_{\rm{H}}}{10^{23}\rm{cm}^{-2}}\right)
\left(\frac{\lambda}{R_{\rm{c}}}\right),
\end{equation}
where $E_{\rm{abs}}$ is the rest energy of the absorption line,
$A_{\rm{Fe}}$ is the abundance of iron
(the solar abundance of iron is $A_{\rm{Fe}\sun}=3.3\times 10^{-5}$,
Morrison \& McCammon 1983), $f_{\rm{l}}$ is the fraction
of ions in the H-like or 
He-like state,
$f_{\rm{lu}}$ is the absorption oscillator strength
and $N_{\rm{H}} (\equiv n_{\rm{H}}R_{\rm{c}})$ 
is the parameter {\it related} to the column density. 
We checked that although for some photon trajectories, energies and adopted 
values of $N_{\rm{H}}$,
the optical depth for resonant absorption may exceed unity, 
the depth of the observed continuum absorption feature varies
almost linearly with $N_{\rm{H}}$ (for different inclinations of the disc) 
implying that we are effectively dealing with
optically-thin resonant absorption. 



\subsection{The iron line profile}
\subsubsection{The absorption of the continuum and the emission line}
The overall iron line profile was calculated in two steps.
In the first step we calculated the shape of the partially absorbed 
fluorescent emission
line generated in the disc and, superimposed on it,   
the resonant absorption feature due to the continuum absorbed by 
the the hot cloud surrounding the disc. 
We assumed that the continuum was generated in the corona located 
just above the accretion disc.
We used the ray back-tracing method and searched
for the absorption regions corresponding to all energies
of the continuum photons and the fluorescent line photons and calculated
the optical depth as a function of energy for each photon trajectory.
We then accumulated the optical depths when multiple
absorption points occurred for the same energy.
The line energy flux relative to the continuum for a given pixel element
on the observer's image plane $F_{\rm{line}}$ was calculated 
from the following formula:
\[
F_{\rm{line}}(E_{\rm{obs}})dE_{\rm{obs}}=
\left[g_{\rm{t}}^{4}\Delta I(E_{\rm{obs}}/g_{\rm{t}}) \exp\left(-\sum\tau\right)EW_{\rm{em}}+\right
.\]
\begin{equation}
\left.-g_{\rm{t}}^{3}I_{\rm{c}}(E_{\rm{obs}}/g_{\rm{t}})\left(1-\exp\left(-\sum \tau\right)\right)
dE_{\rm{obs}}\sec\beta\right]\frac{dx dy}{r_{0}^{2}},
\end{equation}
where $\Delta I(6.4 \rm{keV})=I_{c}(6.4 \rm{keV})$ and $0$ otherwise,
$EW_{\rm{em}}$ is the equivalent
width of the disc emission line from the disc element as seen face on in the 
local rest frame, 
$g_{\rm{t}}$ is the total redshift factor, $\beta$
is the angle between the direction perpendicular to the disc and the direction
of emitted photons as measured by the comoving observer and the last factor 
represents the solid angle subtended by the pixel on the observer's
image plane.
The continuum intensity $I_{\rm{c}}$ in the above equation is given by
$I_{\rm{c}}=\frac{1}{\pi}(\varepsilon (R)/2)$,
where $\varepsilon (R)\propto R^{-\Gamma}E_{\rm{em}}^{-\alpha}$ 
is the total X-ray emissivity of the corona
at the distance $R$ from the centre.  
The second component in the above equation describes absorption of
the continuum photons, whereas the first corresponds to the absorption
of the disc emission line measured relative to the continuum.
The $\sec\beta$ factor accounts for the isotropy of radiation from the 
corona which was assumed to be optically thin.
\subsubsection{Correction for spontaneous emission following absorption}
Resonant absorption
is a scattering process and therefore in our case is
followed by de-excitation of the iron ions rather than destruction 
of photons. This effect will of course partially cancel the absorption 
features. The importance of this effect may be estimated for the
case of a point-like source in the geometrical centre of the disc. 
Such estimation may be reasonable if the
emissivity index of the disc is steep. However at this stage we need this 
assumption only to estimate the order of magnitude of this effect.
In such a situation the emission line flux $F_{\rm{dex}}$ is given by 
(see Matt 1994):
\begin{equation}
F_{\rm{dex}}=0.5F_{\rm{abs}}Y_{i-1}\left(\frac{\Delta\Omega/4\pi}{0.5}\right),
\end{equation}
where $F_{\rm{abs}}$ is the absorbed flux and $Y_{\rm{i}-1}$ is the probability
that the absorption is followed by a
spontaneous emission rather than
autoionization (where appropriate) and thus destruction of the photon. 
This probability can be approximated as the fluorescent
yield of the previous ion (Band et al. 1990) and equals 1 in the case
of H and He-like iron ions.
Taking into account the assumed geometry of the system, $\Delta\Omega=0.5$
(if the disc is cold), the reemitted flux may be as large as
$F_{\rm{dex}}=0.5F_{\rm{abs}}$. Therefore in the subsequent calculations we 
took into account this effect and considered
a fully relativistic treatment of this contribution to the overall 
line profile. Note that the magnitude of this correction
may be overestimated in the Newtonian approximation. This is because
in the real situation spontaneous emission from  
ions is Doppler boosted in the 
direction almost perpendicular to the line-of-sight (for low inclinations
of the accretion disc relative to the observer) and the photons
are preferentially radiated towards the accretion disc because
of the light bending. Note that the above estimate is 
an upper limit also because the reabsorption of the line has been neglected.\\
\indent
We separately included the contribution to
the iron line from 
spontaneous emission following absorption 
by taking into account
the photons from the disc and the corona which were initially 
radiated in some other direction than that to the observer 
and later re-emitted into the line-of-sight. This was done by forward
integrating the isotropic distribution of a large number of 
continuum and line photons from the whole corona located just
above the disc and the whole disc respectively. 
For each trajectory and initial rest energy we localized the absorption
regions and accumulated the absorbed luminosity
in a large number of zones within the cloud. We assumed that either 
photons left the cloud without scattering or were scattered
only once, i.e. we neglected the reabsorption of the line photons. 
This is a good approximation if the plasma is optically 
thin in the line. 
Note also that spontaneous emission which follows absorption produces photons
at the energy of resonant transition and such photons are less likely to be
further resonantly absorbed because of large energy shifts as seen by the local
observers in the plasma. In other words, observers at some region in the 
absorbing plasma may see the re-emitted photons from other parts of the cloud 
at energies different from this of resonant transition and therefore such 
photons will not be reabsorbed.
The total intercepted luminosity in any given zone within the cloud 
was calculated
from (see also Ruszkowski 1999):
\begin{equation}
L(r,\theta)=\int\frac{\varepsilon (R)dS}{N}\sum_{\rm{i}=1}^{N}Q(R,r,\theta) ,
\end{equation}
where the summation is over all photon trajectories and emission energies,
$N$ is the total number of trajectories and
$dS$ is the surface area of an element of the corona as measured by the
corotating observer:
\begin{equation}
dS=\gamma^{-1}\left(\frac{A}{\Delta}\right)^{1/2}\left(1+
\frac{2aV^{(\phi)}}{r\Delta^{1/2}}\right)^{-1}d\phi dr ,
\end{equation}
where $\gamma$ and $V^{(\phi)}$ are the Lorentz factor and the azimuthal
component of the velocity of the relative motion of the corona
and the locally non-rotating observer. The remaining symbols have
their usual meaning in the Kerr metric. The factor $Q(R,r,\theta)$ 
in Eq. 9 is:
\begin{equation}
Q(R,r,\theta)=\left\{ \begin{array}{ll}
g_{\rm{m}}^{2}P_{\rm{m}}EW_{\rm{em}}\cos \beta & \mbox{for the line}\\
g_{\rm{m}}^{2}P_{\rm{m}}dE_{\rm{em}} & \mbox{for the continuum}\\
                    \end{array}
\right. 
\end{equation}
where m is the number of an absorption region for a given trajectory
and emission energy and 
\begin{equation}
P_{\rm{m}}=(1-e^{-\tau_{\rm{m}}})\prod_{\rm{i}=0}^{\rm{m}-1}e^{-\tau_{\rm{i}}}.
\end{equation}
The emission from the zones in the ionized cloud
was then treated separately 
and each zone served as a source of line radiation.
We calculated the observed 
flux by following a bundle of photons from a particular position within
the cloud and computing the distortions of the solid angle 
$\Delta\Omega_{\rm{em}}$ subtended by the surface element $\Delta S_{\rm{obs}}$
on the observer's image plane as measured
in the frame of reference of the emitter moving with the plasma flow.
We used a formula similar to (4), i.e. :
\begin{equation}
F_{\rm{dex}}(E_{\rm{obs}})dE_{\rm{obs}}=g^{2}\frac{L(r,\theta)}{4\pi}
\left.\frac{\partial\Omega_{\rm{em}}}{\partial S_{\rm{obs}}}\right|_{\rm{i}},
\end{equation}
where the derivative was calculated at a fixed inclination $i$ 
of the disc relative to the observer.

In order to obtain the overall line profile we
added the fluxes calculated from Eq. 7 and Eq. 13.
We also computed the observed equivalent widths 
of the lines by
integrating the obtained profiles divided by the observed continuum.

\section{Results and Discussion}
The examples of the computed iron line profiles for the 
parameters similar to that obtained by Nandra et al. (1999) 
for NGC 3516 are shown in Fig. 2 and Fig. 3. 
\begin{figure}
\centerline{\psfig{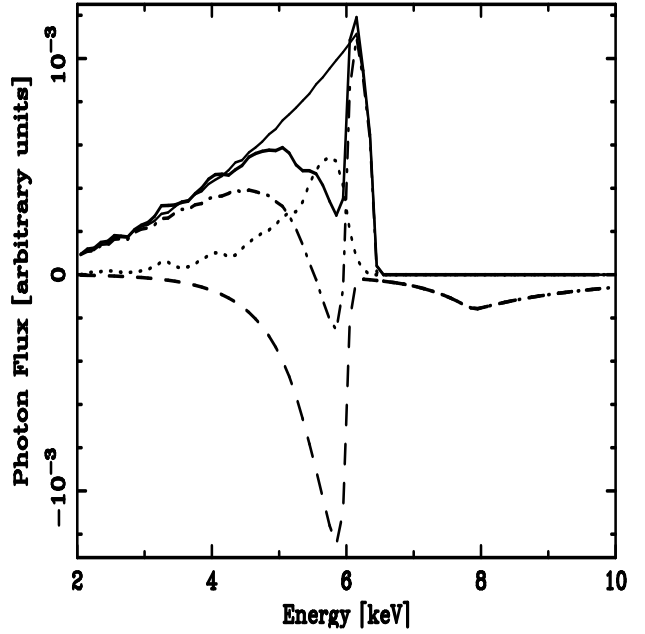}}
\caption{The iron line profile for: $i=10^{\rm{o}}$, $EW_{em}=782$ eV,
$\widetilde{EW}_{\rm{obs}}=510$ eV (unobscured; i.e disc emission line only),
$EW_{\rm{obs}}=398$ eV (obscured), 
$\Gamma=3.0$, $\alpha=0.5$, $r_{\rm{out}}=100$m (outer radius of the disc,
$R_{\rm{c}}=10$m, $N_{\rm{H}}=4\times 10^{23}$ cm$^{-2}$, a=0.998, 
$A_{\rm{Fe}}=2 A_{\rm{Fe}\odot}$, absorption by He-like Fe ions; 
thin solid line - disc fluorescent line, thick solid line - resonantly
absorbed disc fluorescent line, dashed line - resonant absorption of the 
continuum, dash-dotted line - disc fluorescent line absorbed by resonant and 
photoelectric absorption, dotted line - correction for spontaneous emission 
following absorption}
\end{figure}
\begin{figure}
\centerline{\psfig{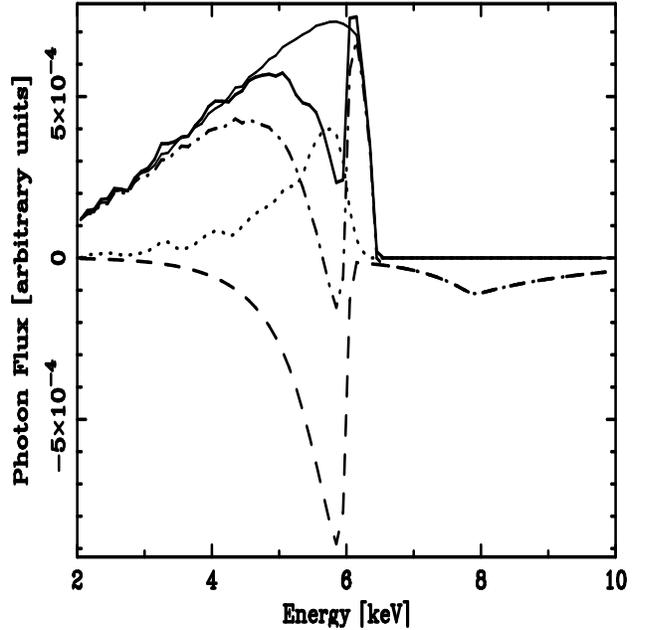}}
\caption{As Fig. 1 but for $EW_{\rm{em}}=1168$ eV, $\widetilde{EW}_{\rm{obs}}=698$ eV, $EW_{\rm{obs}}=578$ eV, $\Gamma=3.25$}
\end{figure}
(Note that the exact values of the observed 
line equivalent widths are not known
because the iron line in NGC 3516 varies on short time scales and only 
time averaged equivalent width is reported in Nandra et al. 1999.)
The thick red line denotes the total iron line
profile (photon flux vs. energy). 
The shape of the line on Fig. 2 may provide an acceptable fit 
to the temporal profile obtained by Nandra et al. (1999) 
(compare with the inset in
their Fig. 1). It peaks around 6.2 keV, has a long redshifted tail 
and possesses the redshifted absorption feature below 6.0 keV. 
As can be seen on Fig. 2, the observed redshift of the absorption
feature can be mostly accounted for by the 
the strong gravitational redshift alone.
This has to be contrasted with the main interpretation originally given by 
Nandra et al. (1999), who suggest that the redshift of the resonant 
absorption feature may be due to the scattering of radiation by matter
infalling onto the black hole. 
Such matter would have to infall along the black hole rotation axis
in order to efficiently absorb radiation from the central parts of the disc.
Thus their interpretation implicitly implies accretion with low angular 
momentum which is less likely. Therefore our results suggest that the 
current observational data do not necessarily give the first 
direct evidence for the accretion of matter in AGNs.\\
\indent
The other profiles seen on the Fig. 2 and Fig. 3 correspond to different
contributions to the total line profile, namely: 
disc fluorescent emission (thin solid line), resonant absorption of the 
continuum
(dashed line) and correction for spontaneous emission from ions which 
follows absorption (dotted line).
The dash-dotted line shows the disc line absorbed by resonant 
and photoelectric absorption (a small iron absorption edge created 
in the cloud is visible around 8 keV). 
Note that there is no intrinsic resonant absorption of the
fluorescent disc emission line and only the continuum radiation is resonantly
absorbed. This is the result of the low inclination of the disc. Emission line
photons with rest energies at 6.4 keV leave the disc and continually
move towards
increasing gravitational potential and therefore their locally observed 
energies gradually decrease and can never exceed the rest energy of the 
resonant absorption line at 6.7 keV. Note also that the observed equivalent
width of the emission line $\widetilde{EW}_{\rm{obs}}$ (for the unabsorbed
line) is smaller than the 
equivalent width $EW_{\rm{em}}$ 
measured face on in the local rest frame because 
of the gravitational redshift and the light bending which leads to the 
central parts of the disc being effectively 
observed at large angle even though the disc is seen at low inclination.
The characteristic skewed shape of the profile of 
resonant absorption of the continuum
is the result of the systematic effect of the gravitational redshift which
creates the long redshifted tail and also due to variations of the 
Sobolev length across the disc as seen by the observer. The coherence length
is larger for the photons on the approaching side of the disc and smaller 
on the opposite side because the cloud on the receding side 
has a velocity component directed away from the observer 
so the photons and plasma travel locally in opposite directions.
Fig. 2 and Fig. 3 also show the contribution to the 
line from 
spontaneous emission following absorption.
As expected, the profile of this line
has a shape similar to that of the fluorescent disc line.\\
\indent
The line photons emitted by the cloud
which impinge on the cold disc are thermalized and do not contribute to 
the final iron line profile. 
It has to be stressed that, in modelling NGC 3516, we assumed that the 
accretion disc was cold. The large observed equivalent widths of the order of
500 eV may suggest an ionized disc, 
however a cold disc with an overabundance of 
iron and underabundance of other elements may still produce large equivalent 
widths of iron line by fluorescence from the illuminated disc material
(Reynolds et al. 1995, Lee et al. 1999). 
The enhanced abundance of iron in the central
regions of AGN is also plausible if one considers that in quasars
large 
metallicities
of up to 9 times the solar value
have been inferred (Hamann \& Ferland 1999).
If however the accretion disc is ionized,
two processes work in the opposite directions:
1) preferential illumination of the disc and efficient reflection
of the line photons from the cloud 
and 2) Comptonization in the disc which smears this line.
These two processes cancel each other to some extent.
The continuum produced in the corona can also be reflected and later 
absorbed in the cloud which will lead to increases in the strengths of the
continuum level, absorption feature and disc line
by the same factor (for fixed $EW_{\rm{em}}$).
Note that even when the disc is ionized,
the albedo may still be significantly far from unity. Thus,
for an ionized accretion disc, 
resonant absorption is not completely cancelled
by the correction for spontaneous emission.\\
\indent
The depth of the absorption feature depends on our parameter 
$N_{\rm{H}}$ which
is related to the column density. The amount of ionized plasma which
reprocesses the continuum and the emission line from the disc is constrained
by the nondetection of the absorption edge of iron in the profiles obtained
by Nandra et al. (1999). 
In order to estimate the importance of the absorption edge we calculate
the factor $1-e^{-\tau_{\rm{edge}}}\approx\tau_{\rm{edge}}$ 
where $\tau_{\rm{edge}}$ is the optical depth
at the photoionization threshold energy:
\[
1-e^{-\tau_{\rm{edge}}}=
\]
\begin{equation}
\hspace{1cm}0.12\left(\frac{f_{\rm{l}}N_{\rm{H}}}{10^{23}\rm{cm}^{-2}}\right)
\left(\frac{\sigma_{\rm{egde}}}{2\times 10^{-20}\rm{cm}^{2}}\right)
\left(\frac{A_{\rm{Fe}}}{2A_{\rm{Fe}\sun}}\right) ,
\end{equation}
where $\sigma_{\rm{edge}}$ is the cross section for
photionization 
($\sigma_{\rm{H}}$ and $\sigma_{\rm{He}}$ are both close 
to $2\times 10^{-20}\rm{cm}^{2}$,
Yakovlev et al. 1992 and references therein).
In consequence, the contrast 
of the absorption edge relative to the continuum can be small.
The definition of $N_{\rm{H}}$ does not include the fact that the 
column densities for geodesics intersecting the cloud further away from the 
centre are smaller, which will also reduce the iron edge.
It also has to be stressed that
the contrast of 
any absorption edge created close to the black hole would be 
further reduced
by a factor of a few by the strong relativistic effects operating
in this region. 
We therefore calculate the 
iron edge using a fully relativistic treatment to determine 
the allowed values of $N_{\rm{H}}$. The results are shown
on Fig. 2 and Fig. 3 (dash-dotted lines around 8 keV). 
The strength of the calculated 
absorption edge is well within the acceptable range given the quality of 
the current data.
In the considered range of temperatures the electrons in the plasma are 
nonrelativistic or mildly relativistic 
and the Thomson optical depth is 
$\tau_{\rm{Th}}\approx 6.6\times 10^{-2}(N_{\rm{H}}/10^{23}\rm{cm}^{-2})$
where $N_{\rm{H}}$ is the hydrogen column density.
This means that the Compton $y$ parameter
$y=4\tau_{\rm{Th}}(1+\tau_{\rm{Th}})\Theta (1+\Theta)\ll 1$, where 
$\Theta=kT/mc^{2}=1.7\times 10^{-2}(T/10^{8}\rm{K})\ll 1$
and thus, for the adopted values of $N_{\rm{H}}$,
Comptonization by the Thomson thin plasma 
is unlikely to significantly alter the energies of the emission line
photons radiated by
the accretion disc and the continuum in the range $2 - 10$ keV
(see also Fabian et al. 1995).
As discussed above, the plasma may be photoionized by the continuum radiation,
so in addition to the above processes,
fluorescence following photoionization in the cloud 
should in principle be considered.
Its influence will however be small.
It can be roughly estimated by calculating the ratio of 
the equivalent width for the fluorescent photons $EW_{\rm{f}}$ to the 
equivalent width for the resonant absorption $EW_{\rm{r}}$ 
in the Newtonian limit.
In our model the plasma is effectively optically thin and in 
this regime the ratio $EW_{\rm{f}}/EW_{\rm{r}}$ 
can be calculated by using the results
from Matt (1994) and Krolik \& Kallman (1987) and cast in the 
following form: 
\[
\frac{EW_{\rm{f}}}{EW_{\rm{r}}}=0.14
\left(\frac{\sigma_{\rm{edge}}}{2.0\times10^{-20}\rm{cm}^{2}}\right)
\left(\frac{\langle Y\rangle}{0.5 f_{\rm{l}}}\right)
\left(\frac{4}{3+\alpha}\right)\times
\]
\begin{equation}
\hspace*{1cm}\left(\frac{(E/E_{\rm{edge}})^{\alpha}}{0.9}\right)
\left(\frac{E}{6.7 \rm{keV}}\right)
\left(\frac{\Delta\Omega/4\pi}{0.5}\right)
\left(\frac{f_{\rm{lu}}}{0.5}\right)^{-1} ,
\end{equation}
where $\langle Y\rangle$ is the fluorescent yield averaged over the fractional 
abundance of the iron ions
(depending on temperature and some uncertain atomic physics (un the case of 
He-like Fe), $Y_{\rm{H}}$ and $Y_{\rm{He}}$ lie in the range 
$\approx 0.5-0.7$, Krolik 1999).
$\alpha$ is the spectral energy index, 
$E$ is the energy of the absorption/fluorescent line, $E_{\rm{edge}}$ is the
threshold energy for photoionization
and $\Delta\Omega$ is the solid angle subtended by the ionized plasma
as seen by the primary continuum source.
If the accretion disc is cold then the fluorescent photons
created in the hot cloud are thermalized when they
impinge upon the disc surface and are effectively destroyed
and therefore $\Delta\Omega/4\pi=0.5$. Similarly in the case of the hot disc,
fluorescent photons directed towards the disc are strongly Comptonized
by multiple scatterings in the hot, optically thick medium and 
the line spreads out to some extent and thus the effective value of 
$\Delta\Omega/4\pi<1$.
For example, the above ratio of the equivalent widths 
for $\alpha=1$ can be as small as
0.065 in the case of the cold accretion disc surrounded by a hot plasma
in which He-like ions are the primary source of opacity.
Moreover, this is an upper limit because the reabsorption of the fluorescent 
photons has been neglected.\\
\indent
Fig. 4 shows the effect of the size of the cloud on the line profile. 
\begin{figure}
\centerline{\psfig{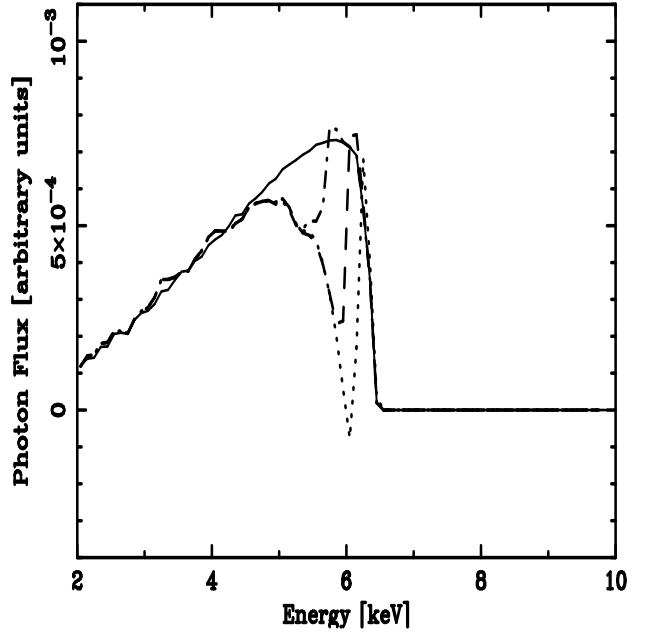}}
\caption{The iron line profiles for the same parameters and the same cloud density as in Fig. 2 
(except for the equivalent widths) but for three different cloud 
sizes $R_{c}=7$m (dash-dotted line), 10m (dashed line), 13m (dotted line);
thin solid line denotes disc fluorescent emission line profile}
\end{figure}
Large cloud sizes lead to stronger absorption because the coherence length
over which resonant absorption can occur increases with distance from 
the black hole. Of course the smaller the cloud the greater the
redshift of the absorption feature, but 
the magnitude of absorption decreases as the size of the cloud
gets smaller. In principle, an ensemble of very small clouds or filaments
distributed at a range of distances from the central black hole could produce
a number of narrow absorption features 
superimposed on the emission line, an effect
similar to the 'Ly$\alpha$ forest' observed in the spectra of distant quasars.
It is indeed plausible that the absorbing plasma may exist in the form of 
small clumps. If the clumpy plasma of smoothed out mean density 
$n_{\rm{H}}$ is in photoionization equilibrium,
the ionization parameter necessary to produce a high ionization state
has to be $\xi=L_{\rm{X}}/(\frac{n_{\rm{H}}}{f} R^{2}_{\rm{c}}) \sim 10^{4}$,
where $f$ is the filling factor. 
Assuming the X-ray luminosity (in the energy range 2-50 keV), 
the size of the absorbing region and the column density
to be $L_{\rm{X}}\sim 10^{43}$ erg s$^{-1}$ (Stripe et al. 1998),
$R_{\rm{c}}\sim 10^{14}$ cm and $N_{\rm{H}}\sim10^{24}$ cm$^{-2}$ respectively,
one can obtain a rough estimate of the filling factor 
$f\propto\xi N_{\rm{H}}R_{\rm{c}}/L_{\rm{X}}\sim 0.1$, implying
that the plasma has a clumpy structure.
We note that the free-free emission from such clumps will
be $L_{\rm{ff}}\propto 10^{40}T_{7}^{1/2}$erg s$^{-1}$, 
much less that the luminosity 
in the X-ray band. 
Magnetic fields are a plausible mechanism of confinement of the clouds
and we note that the existence of magnetically confined clumps
above the accretion disc was proposed and extensively discussed by 
Kuncic, Celloti \& Rees (1997). 
The cloud densities envisaged in our work are however much less extreme.\\
\indent
Recently Iwasawa et al. (1999) reported on the observations of 
the broad iron emission line
in MCG-6-15-30. They observed a bright flare in the light curve during which
the line peaked at 5 keV and most of the line emission was shifted below
6 keV with no component detected at 6.4 keV. They interpreted it as the 
result of an extraordinarily large gravitational redshift owing to a dominant
flare occurring very close to the black hole at $r\approx 2$m. We speculate
that their profile may also be explained by resonant absorption by the 
plasma expelled during the bright flare. 
It is conceivable that such highly ionized material may be ejected to larger 
heights above the accretion disc where the resonant absorption will be more
efficient due to the larger coherence length. The weaker gravitational field
at such distances will then lead to the absorption feature being
less redshifted which may in turn cancel  
the main peak of the disc fluorescent emission line. 
Therefore the maximum of the profile of the partially 
absorbed iron line will be effectively seen at lower energies.
\section{Conclusions}
We have demonstrated that resonant absorption potentially plays an 
important role in the interpretation of the fluorescent iron emission 
line profiles from accreting black holes. Our model can explain
the absorption features seen in the profiles recently 
obtained by Nandra et al. (1999) giving an alternative explanation 
of the data whereby the redshift of the absorption features is mainly 
gravitational in origin and fast accretion of the matter along the spin axis
of the black hole is not required. We hope that forthcoming data from
Chandra X-ray Observatory may help to break the degeneracy of the two proposed 
scenarios. Future high signal-to-noise observations may also provide us
with information about the accretion flow near the black hole.



\section{acknowledgments}
MR acknowledges support from an External Research Studentship of Trinity
College, Cambridge; an ORS award; and the Stefan Batory Foundation. 
ACF thanks Royal Society for support. We thank Giorgio Matt, Roger Blandford
and Martin Rees for useful discussions. We also thank the referee - 
H. Netzer for constructive comments.

\begin{thebibliography}{99}
\bibitem []{} Band D.L, Klein R.I., Castor J.I., Nash J.K.,1990, ApJ, 90
\bibitem []{} Castor J.I., 1970, MNRAS, 149, 111
\bibitem []{} Dabrowski Y., Fabian A.C., Iwasawa K., Lasenby A.N., 
Reynolds C.S., 1997, MNRAS, 288, L11
\bibitem []{} Fabian A.C., Kunieda H., Inoue S., Matsuoka M.,
Mihara T., Miyamoto S., Otani C., Ricker G., Yamauchi M., Yaqoob T.,
 1994, PASJ, 46, L59
\bibitem []{} Fabian A.C., Nandra K., Reynolds C.S., Brandt W.N., 
Otani C., Tanaka Y., Inoue H., Iwasawa K., 1995, MNRAS, 277, L11
\bibitem []{} Hamann F., Ferland G.J., 1999, to appear in ARA\&A and 
available at astro-ph/9904223
\bibitem []{} Iwasawa K., Fabian A.C., Reynolds C.S., Nandra K, Otani C., 
Inoue H., Hayashida K., Brandt W.N., Dotani T., Kunieda H., Matsuoka M., 
Tanaka Y., 1996, MNRAS, 282, 1038  
\bibitem []{} Iwasawa K., Fabian A.C., Young A.J., Inoue H., Matsumoto C.,
1999, MNRAS, 306, 19
\bibitem []{} Kato T., 1976, ApJS, 30, 397
\bibitem []{} Kuncic Z., Celotti A., Rees M.J., 1997, MNRAS, 717, 730
\bibitem []{} Kurpiewski A., Jaroszy\'nski M., 1999, A\&A, 364, 713
\bibitem []{} Krolik J.H., 1999, Active Galactic Nuclei: from the Central 
Black Hole to the Galactic Environment (Princeton: Princeton University Press)
\bibitem []{} Krolik J.H., Kallman T.R., 1987, ApJL, 320, 5
\bibitem []{} Lee J.C., Brandt W.N., Fabian A.C., Iwasawa K., Reynolds C.S,
in preparation
\bibitem []{} Lee J.C., Fabian A.C., Brandt W.N., Reynolds C.S., Iwasawa K,
 1999, MNRAS, in press
\bibitem []{} Matt G., 1994, MNRAS, 267, L17
\bibitem []{} Morrison R., McCammon D., 1983, ApJ, 270, 119
\bibitem []{} Mushotzky R.F., Fabian A.C., Iwasawa K., Kunieda H., 
Matsuoka M., Nandra K., Turner Y., 1995, MNRAS, 272, L9
\bibitem []{} Nandra K., George I.M., Mushotzky R.F., Turner T.J., 
Yaqoob T., 1997, ApJ, 477, 602
\bibitem []{} Nandra K., George I.M., Mushotzky R.F., Turner T.J.,
Yaqoob T., 1999, ApJL, in press, astro-ph/9907193
\bibitem []{} Novikov I.D., Thorne K.S., 1993, in Black Holes 
[Les Houches Summer School 1972], DeWitt C. ed. (New York: Gordon and
Breach Science Publishers) 
\bibitem []{} Reynolds C.S., 1997, MNRAS, 286, 513
\bibitem []{} Reynolds C.S., Fabian A.C., Inoue H., 1995, MNRAS, 276, 1311
\bibitem []{} Shu F.H., Physics of Astrophysics, Volume I: Radiation,
1991 (Mill Valley: University Science Books)
\bibitem []{} Stripe G.M., Wilkes B.J., Comastri A., Mathur S., O'Brien P.T.,
1998, in Scarsi H., Giommi P., Fiore F., ed., The Active X-Ray Sky: Results 
from BeppoSAX and RXTE, Nuclear Physics B Proceedings Supplements, 69, 505
\bibitem []{} Tanaka Y. et al., 1995, Nature, 375, 659 
\bibitem []{} Yakovlev D.G., Band I.M., Trzhaskovskaya M.B., Verner D.A.,
ESO scientific preprint No. 835
\end{thebibliography}
\end{document}